\documentclass[english,aps,prl,reprint,twocolumn,superscriptaddress]{revtex4-1}
\usepackage{lmodern}

\usepackage[T1]{fontenc}
\usepackage[latin9]{inputenc}
\usepackage{babel}
\usepackage{units}
\usepackage{textcomp}
\usepackage{amstext}
\usepackage{amssymb}
\usepackage{graphicx}
\usepackage{subscript}
\usepackage[unicode=true,
 bookmarks=true,bookmarksnumbered=false,bookmarksopen=false,
 breaklinks=false,pdfborder={0 0 0},backref=false,colorlinks=false]
 {hyperref}
\hypersetup{pdftitle={Breaking of symmetry in graphene growth on metal substrates},
 pdfauthor={Artyukhov et al.},
 pdftex,colorlinks=true,urlcolor=blue,linkcolor=blue,citecolor=blue}

\makeatletter

\@ifundefined{textcolor}{}
{%
 \definecolor{BLACK}{gray}{0}
 \definecolor{WHITE}{gray}{1}
 \definecolor{RED}{rgb}{1,0,0}
 \definecolor{GREEN}{rgb}{0,1,0}
 \definecolor{BLUE}{rgb}{0,0,1}
 \definecolor{CYAN}{cmyk}{1,0,0,0}
 \definecolor{MAGENTA}{cmyk}{0,1,0,0}
 \definecolor{YELLOW}{cmyk}{0,0,1,0}
}

\renewcommand{\fnum@figure}{\textbf{FIG. \thefigure}}

\usepackage{rotating}
\usepackage{times,mathptmx} 
\usepackage{microtype}

\makeatother

\begin{document}

\title{Breaking of symmetry in graphene growth on metal substrates}

\author{Vasilii I. Artyukhov}

\affiliation{Department of Materials Science and NanoEngineering, Rice University,
Houston, TX 77005, USA}

\author{Yufeng Hao}

\affiliation{Department of Mechanical Engineering, Columbia University, New York, New York 10027, USA}

\author{Rodney S. Ruoff}

\affiliation{Center for Multidimensional Carbon Materials, Institute for Basic Science (IBS), Ulsan 689-798, Republic of Korea}

\affiliation{Department of Chemistry, Ulsan National Institute of Science and Technology (UNIST), Ulsan 689-798, Republic of Korea}

\author{Boris I. Yakobson}

\email{biy@rice.edu}

\affiliation{Department of Materials Science and NanoEngineering, Rice University,
Houston, TX 77005, USA}
\begin{abstract}
In graphene growth, island symmetry can become lower than the intrinsic
symmetries of both graphene and the substrate. First-principles calculations
and Monte Carlo modeling explain the shapes observed in our experiments
and earlier studies for various metal surface symmetries. For equilibrium
shape, edge energy variations $\delta E$ manifest in distorted hexagons
with different ground-state edge structures. In growth or nucleation,
energy variation enters exponentially as $\sim e^{\delta E/k_{\text{B}}T}$,
strongly amplifying the symmetry breaking, up to completely changing
the shapes to triangular, ribbon-like, or rhombic.
\end{abstract}
\maketitle
While exfoliation techniques can produce monolayers of graphene \cite{2004novoselovelectric}
and other two-dimensional (2D) materials \cite{2011colemantwodimensional}
of extraordinary quality \cite{2008bolotinultrahigh}, their lack
of scalability hampers their use in applications. Chemical vapor deposition
(CVD) synthesis \cite{2014tetlowgrowthof} can address the scalability
concern. However it is difficult to produce graphene samples of quality
comparable to exfoliated layers \cite{2014yazyevpolycrystalline},
motivating both empirical and theoretical effort to understand and
improve graphene growth.

Because CVD involves a solid substrate in contact with graphene, their
interaction alters the latter\textquoteright{}s properties. This influence
cannot be described simply as the interaction of complete graphene
crystal with the support. Instead, the relevant processes occur as
graphene assembles. Due to the inherent difficulty of observing growth
\emph{in situ}, theoretical understanding is indispensable.\medskip

Previous study of the morphology of graphene under kinetic or thermodynamic
control from atomistic level \cite{2012artyukhovequilibrium} was
able to predict many observed shapes such as zigzag-edged (slowest-growing)
hexagons \cite{2011yucontrol,2012genguniform,2013haotherole} or dodecagons
with $19.1^{\circ}$ (fastest-etching) \cite{2013maedgecontrolled}
and $30^{\circ}$ (equilibrium shape) angles \cite{2014chennearequilibrium}.
All these shapes inherit the hexagonal symmetry of graphene. Yet,
recurring observations of less symmetric shapes call for a deeper
study of the effects of the substrate on the growth of graphene. In
this work we reveal how symmetry breaking manifests in graphene growth
and results in shapes with lowered (threefold, twofold) symmetry,
using Ni and Cu substrates as examples. For the equilibrium shape
of graphene on the Ni(111) surface, we show using first principles
calculations how tangential `sliding' breaks inversion symmetry and
leads to different atomistic structures at opposite edges of graphene
islands, yet the effect on the Wulff shape is rather weak. However,
since the growth rates contain the energy terms affected by symmetry
in the exponent, we find that under kinetic control, the asymmetry
is amplified, causing a qualitative transition from hexagonal (equilibrium)
to triangular (growth) shapes. Nucleation statistics, also exponential
in the symmetry-breaking strength, can cause strong selection of just
one of the two near-degenerate stacking `phases' of graphene. Casting
the atomistic insight into a coarse-grained Monte Carlo (MC) model
of growth, we explain our observations of broken-symmetry islands
on different surfaces of polycrystalline Cu foil.

\begin{figure*}
\hfill\includegraphics[width=0.25\textwidth]{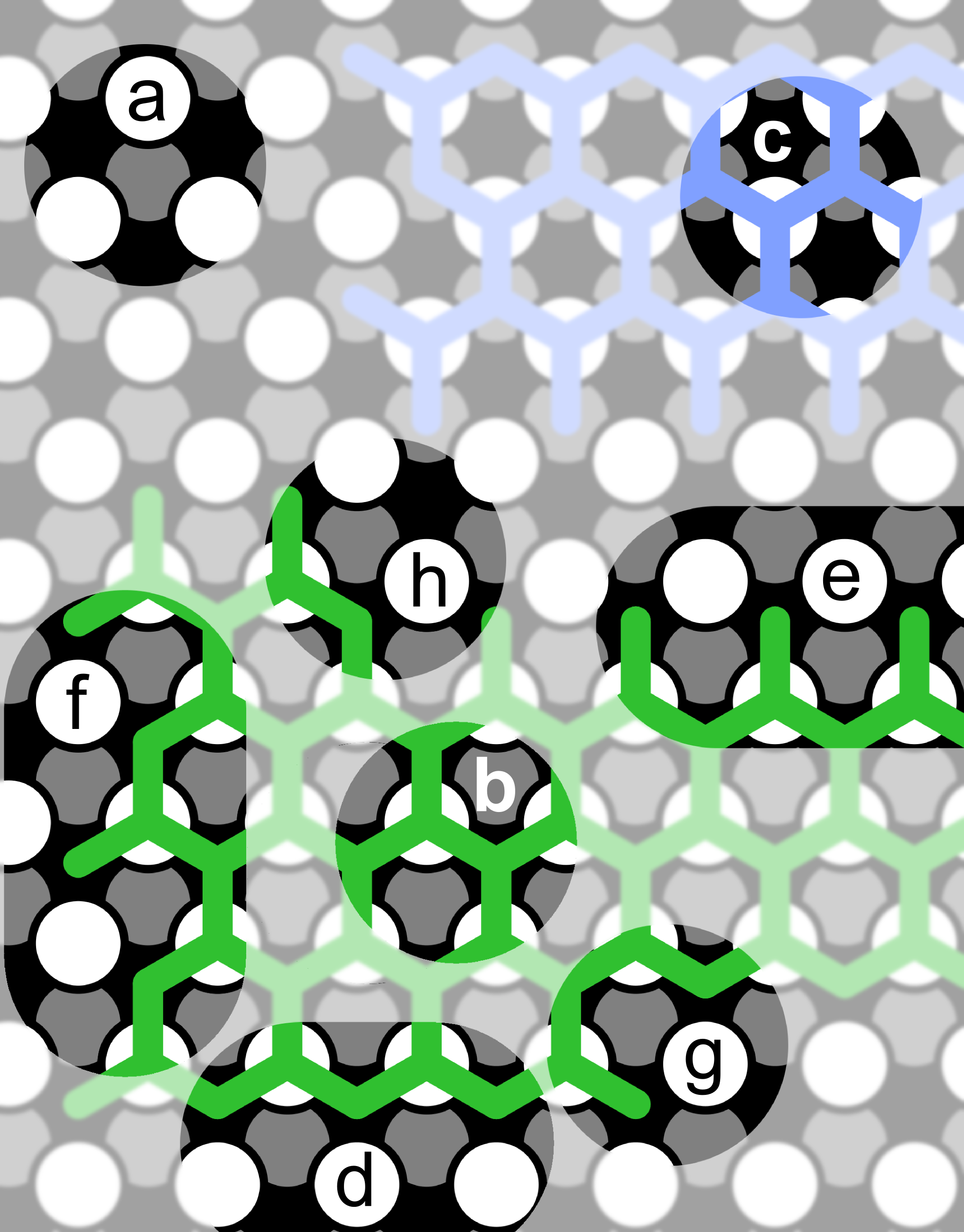} \includegraphics[width=0.65\textwidth]{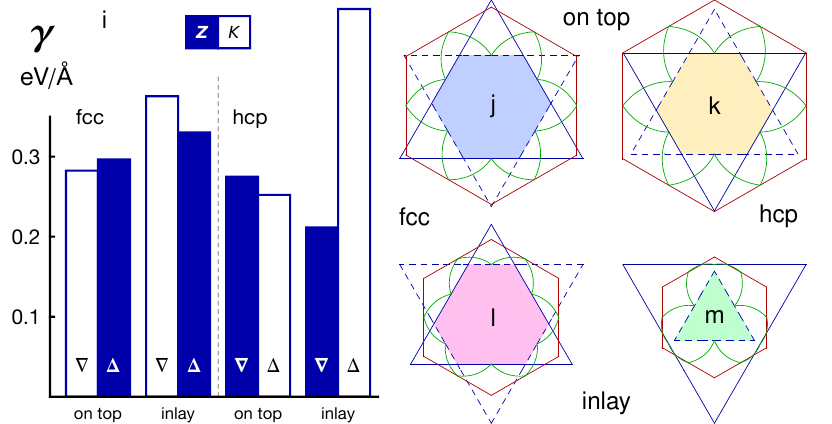}\hfill~

\caption{(\textbf{a--h}) Stacking of graphene on Ni(111) surface. The substrate
\textbf{(a)} is shown in white--gray--black according to the depth.
Graphene is shown in green \textbf{(b)} for \emph{fcc} and blue \textbf{(c)}
for \emph{hcp} stacking. Graphene edges: \textbf{(d)} hexagonal zigzag,
\emph{Z}; \textbf{(e)} Klein, \emph{K}; and \textbf{(f)} armchair
(showing the A5' reconstruction \cite{2012artyukhovequilibrium}).
Edge kinks, \emph{k}, are shown on \emph{Z} \textbf{(g)} and \emph{K}
\textbf{(h)} edges.\textbf{ (i)} Zigzag edge energies computed with
DFT for different stacking (\emph{fcc}, \emph{hcp}), direction ($\nabla,\,\Delta$),
and edge structure (\emph{Z}, \emph{K}) in the on-top and inlay graphene
arrangement with respect to the top Ni layer. The bar width is 0.04
eV/$\text{\AA}$ or 0.1 eV per edge unit cell, corresponding to $\sim k_{\text{B}}T$
at typical growth temperatures ($\sim1000$ K). \textbf{(j--m)} Wulff
constructions for the respective cases: (green) edge energy $\gamma\left(\chi\right)$;
blue lines denote (solid) \emph{Z} and (dashed) \emph{K} edges.\label{fig:Wulff}}
\end{figure*}

Typically, graphene is incommensurate with substrates used for CVD
growth. Without translational invariance even the most basic concepts
such as interface energy and Wulff construction cease to be a reliable
foothold. Therefore we first focus on the important special case of
Ni(111) substrate (with Co(0001) being essentially analogous) where
graphene can stretch by \textasciitilde{}1\% to accommodate the lattice
constant of metal surface, resulting in perfect epitaxial matching.
Since both the (111) surface and graphene have a sixfold rotation
axis, it is possible to form an interface that preserves this symmetry.
However it turns out to be unstable with respect to tangential displacements
that break the alignment of the $C_{6}$ axes producing other stacking
phases which reduce the symmetry to threefold (or even twofold). Typically
one sublattice of carbon atoms is on top of upper-layer Ni atoms,
and the other either in \emph{fcc} of \emph{hcp} sites of the Ni lattice,
forming two almost-degenerate structures (\textbf{Fig. \ref{fig:Wulff}})
that differ only in positioning with respect to the 2\textsuperscript{nd}
layer of Ni. Either way the overall symmetry is triangular rather
than hexagonal and graphene sublattices become inequivalent---like
in BN \cite{2011liubnwhite}. In particular, the six previously degenerate
\emph{Z} edges split up into two triplets, denoted as `$\nabla$'
and `$\Delta$'.\medskip

We begin with determining the substrate effect on the equilibrium
shape of graphene on Ni(111). The edge energy for arbitrary orientation
$\chi$ can be expressed analytically from basic armchair (\emph{A})
and zigzag (\emph{Z}) edge energies \cite{2010liugraphene}, accounting
for inequivalent \emph{Z} edges: $\gamma\left(\chi\right)=2\gamma_{A}\sin\left(\chi\right)+2\gamma_{i}\sin\left(30^{\circ}-\chi\right)$.
Here $\chi$ is the angle with respect to the closest $\nabla$ direction,
$i=\nabla$ when $\left|\chi\right|$ mod $60^{\circ}<30^{\circ}$,
and $i=\Delta$ otherwise. Since the \emph{A} edge symmetry is not
broken by the substrate, $\gamma_{A}$ is known from previous work
\cite{2012artyukhovequilibrium}, leaving us with just the two zigzag
edge energies $\gamma_{\nabla}$ and $\gamma_{\Delta}$ to compute.
As a consequence of inversion symmetry breaking, not only the energies
of $\nabla$ and $\Delta$ edges can be different, but the edges can
have different ground-state atomistic structures. Using density functional
theory computations \cite{1993kresseabinitio,*1996kresseefficient,1994blochlprojector,*1999kressefromultrasoft,1980ceperleygroundstate,1996perdewgeneralized,*1997perdewgeneralized}
(details in Supplemental Material %
\footnote{See Supplemental Material at http://prl.aps.org for for details of computations and edge energy fitting.%
}), we screened a total of 12 possible combinations: 2 for stacking
(\emph{fcc}, \emph{hcp}) \texttimes{} 2 for direction (\emph{$\nabla,\,\Delta$})
\texttimes{} 3 structures (conventional hexagonal zigzag, \emph{Z};
Klein, \emph{K} \cite{1994kleingraphitic}; pentagon-reconstructed
Klein---e.g. \cite{2013wagnerstablehydrogenated}). The pentagon--Klein
reconstruction is always unfavorable. For both stackings, one of the
ground-state edge structures is \emph{Z} but the opposite side favors
\emph{K} (\emph{fcc}: $K_{\nabla}||Z_{\Delta}$, \emph{hcp}: $Z_{\nabla}||K_{\Delta}$)
as \emph{top} C atoms cannot form in-plane bonds with Ni atoms and
prefer to be three-coordinated. To determine edge energies in the
absence of inversion symmetry, we used a series of increasingly larger
triangular islands with only $\nabla$ or only $\Delta$ edges \cite{2011liubnwhite},
by fitting their energies as $E(N)=aN+b\sqrt{N}$ where $N$ is the
number of atoms ~\cite{Note1}. We considered two scenarios, with
graphene flakes on top of the Ni(111) surface or inlaid in the topmost
Ni plane \cite{2013paterainsitu}. In all cases (\textbf{Fig. \ref{fig:Wulff}i})
the energies of $\nabla$ and $\Delta$ edges are close, except for
the inlay-\emph{hcp} case where the \emph{$\Delta$} direction is
impossible to interface with the Ni lattice without a large geometrical
strain (hence the outstandingly high value).

By plotting $\gamma\left(\chi\right)$ in polar coordinates \cite{1951herringsometheorems}
(green line in \textbf{Fig. \ref{fig:Wulff} (j--m)}) we obtain the
Wulff construction. We find that the equilibrium shapes are truncated
triangles (lowered-symmetry hexagons), except for the case of inlay-\emph{hcp}
stacking. Truncated shapes were indeed observed on Ni(111) while this
manuscript was in preparation \cite{2014garcia-lekuespindependent},
and according to our calculations these islands should have Klein
edges in three out of six directions. Yet multiple other observations
show sharp-cornered triangles on top of metal surface \cite{2012olleyieldand,2013paterainsitu},
impossible to explain thermodynamically (\textbf{Fig. \ref{fig:Wulff}}).
This compels us to investigate growth kinetics.\medskip

\begin{figure}
\includegraphics[width=1\columnwidth]{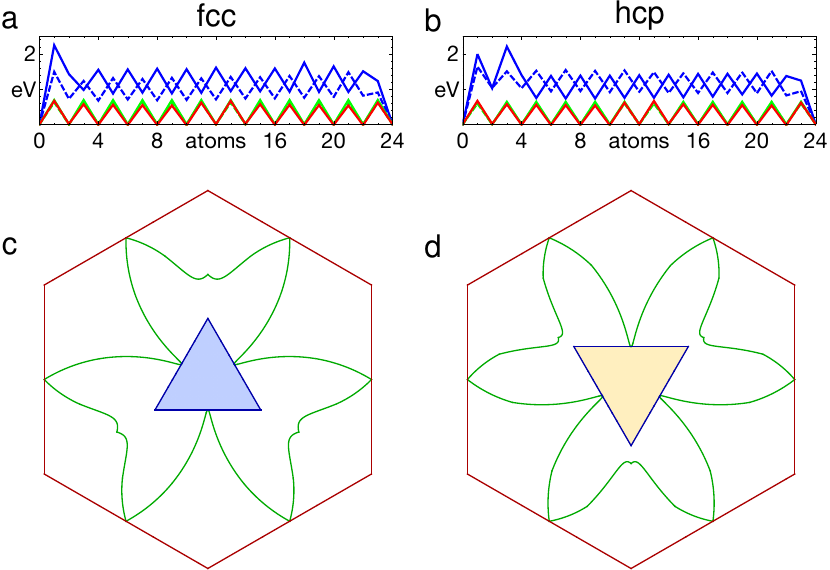}

\caption{\textbf{(a, b)} Free energy evolution during graphene edge growth
in \textbf{(a)} \emph{fcc} and \textbf{(b)} \emph{hcp} stacking: (blue,
solid) \emph{Z}, (blue, dashed) \emph{K}, (green) intermediate, and
(red) armchair edges. \textbf{(c, d)} Kinematic Wulff constructions
for \textbf{(c)} \emph{fcc} and \textbf{(d)} \emph{hcp} stackings:
(green) polar plot of edge growth velocity, (red) velocity of armchair
edges, (blue) velocity of \emph{Z} edges. \emph{K} edges are absent
from the construction. The temperature is set high (0.3 eV) in order
to `compress' the plots in the radial dimension.\label{fig:kWulff}}
\end{figure}

Our ``nanoreactor'' model of graphene growth \cite{2012artyukhovequilibrium}
is naturally extendable to the case of inversion-inequivalent zigzag
edges. We consider only the ground-state edge structures in the on-top
scenario. Carbon atoms are added sequentially to the edges (\textbf{Fig.
S1}~\cite{Note1}) obtaining the free energy sequences shown in \textbf{Fig.
\ref{fig:kWulff}} for (\textbf{a}) \emph{fcc} and (\textbf{b}) \emph{hcp}
stacking. The familiar `nucleation--kink flow' picture is clearly
observed in this plot \cite{2012artyukhovequilibrium}. In either
stacking, the hexagonal \emph{Z} edge (blue solid line) has a higher
free energy barrier for the formation of a new atomic row than \emph{K}
(blue dashed line): 2.24 vs. 1.49 eV for \emph{fcc} and 2.20 vs. 1.633
eV for \emph{hcp} (difference $\Delta E\approx0.6\text{--}0.7$~eV).
And because the rate of formation of new atomic rows at the edge depends
on these energy barriers exponentially, \emph{K} edges will grow much
faster than \emph{Z} and disappear from the growth shape. The closed-form
expression for graphene edge growth velocity \cite{2012artyukhovequilibrium}
can be used (with appropriate modifications to account for broken
symmetry) to plot the kinematic Wulff constructions. As seen in \textbf{Fig.
\ref{fig:kWulff} (c,d}), the result is a triangle with \emph{Z} edges,
\emph{$\Delta$} for \emph{fcc} stacking and \emph{$\nabla$} for
\emph{hcp}.\medskip

\begin{figure}
\includegraphics[width=1\columnwidth]{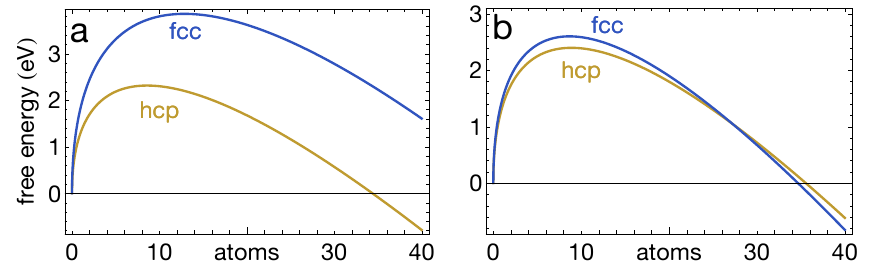}

\caption{Free energy as a function of number of atoms in an island ($\Delta\mu=0.3$
eV) for \textbf{(a)} inlay and \textbf{(b)} on-top scenarios.\label{fig:nucl}}
\end{figure}

The essentially equal values of rate-limiting barriers $E_{Z}$ for
the \emph{hcp} and \emph{fcc} stackings predict similar growth rates.
However, to assess their relative abundance one also needs to consider
nucleation \cite{2014artyukhovwhynanotubes}. Based on the edge energies
and difference between \emph{fcc} and \emph{hcp} 2D bulk energies
($\sim0.03$ eV/atom from the computations) the free energy of an
island can be expressed as a function of its area (number of atoms),
$G=\left(\epsilon-\mu\right)N+c\gamma\sqrt{N}$, where $\epsilon$
is the 2D `bulk' energy of the respective graphene phase, $\mu$ is
the chemical potential, $\gamma$ is the edge energy, and $c$ is
a form factor to discriminate between triangles and hexagons (which
we approximate as perfect).\textbf{ Fig. \ref{fig:nucl}} shows free
energy $G\left(N\right)$ plots for the two `phases' at a chemical
potential bias of $\mu-\epsilon_{fcc}\equiv\Delta\mu=0.3$ eV for
the \textbf{(a)} inlay and \textbf{(b)} on-top scenarios. In the former
scenario \textbf{(a)}, despite the triangular shape of \emph{hcp}
domains, low edge energy yields a much lower nucleation barrier---by
1.54 eV in this example. This leads to a nucleation rate difference
$e^{\left(G_{hcp}^{*}-G_{fcc}^{*}\right)/k_{\text{B}}T}\sim10^{6}$
in favor of the higher-energy \emph{hcp} phase. Here again exponentiation
greatly amplifies the symmetry-breaking effect (compared to the ratio
of $\gamma$ which is only $\sim1.5$). While the growth shapes of
the two phases are oppositely oriented (\textbf{Fig. \ref{fig:kWulff}
(c,d)}), selective nucleation eliminates one of the two possibilities.
This explains the recent observations of co-oriented graphene triangles
on Ni(111) \cite{2013liequilibrium}.

In contrast, for graphene islands on top of the surface the nucleation
barrier difference is merely on the order of $k_{\text{B}}T$ \textbf{(b)},
implying weak if any selectivity (the \emph{hcp}--\emph{fcc} preference
is reversed around $\Delta\mu=0.18$ eV). Indeed, both phases were
identified via characteristic `translational grain boundary' defects
\cite{2010lahirianextended} and by direct observations \cite{2014bianchiniatomicscale}.
\medskip

\begin{figure}
\includegraphics[width=1\columnwidth]{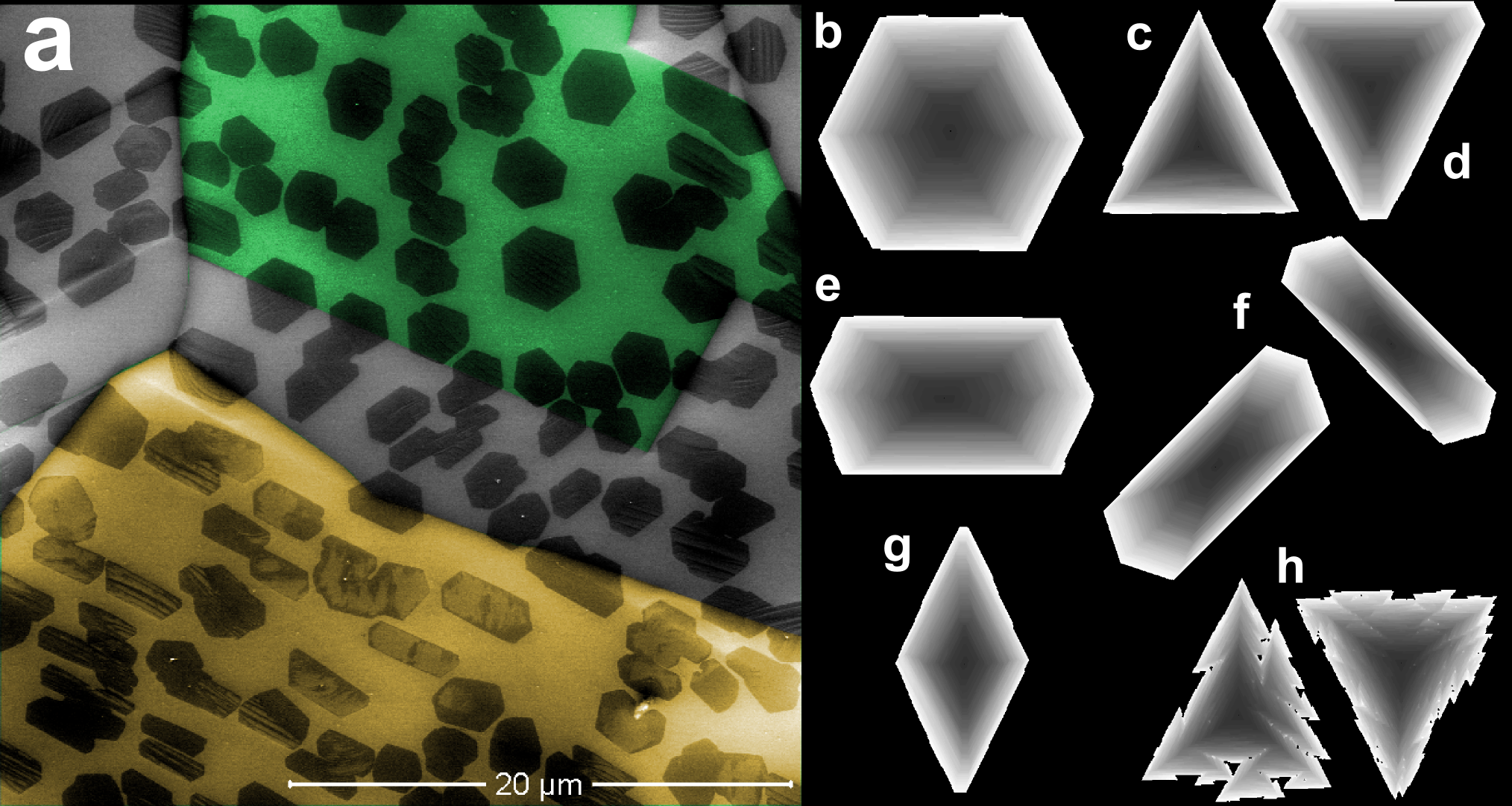}

\caption{\textbf{(a)} Scanning electron microscopy image of graphene grains
on a polycrystalline Cu foil with two largest grains highlighted in
color. \textbf{(b--h)} Monte Carlo modeling of growth: \textbf{(b)}
isotropic kinetics, Cu(111); \textbf{(c)} triangle with a $10^{3}$
($7\, k_{\text{B}}T$) difference between growth probabilities for
\emph{$\nabla$} and \emph{$\Delta$} directions, representing Ni(111);
\textbf{(d)} triangle with a $10^{1}$ ($2\, k_{\text{B}}T$) \emph{$\Delta:\nabla$}
probability ratio; \textbf{(e)} two slow directions, representing
the rectangular Cu(110) surface; \textbf{(f)} two slow directions
with two degenerate orientations, Cu(100); \textbf{(g)} same as \textbf{(e)}
but with two fast directions; \textbf{(h)} calculations with diffusion.
All simulations were run for 30 000 steps. Brightness represents time
(lighter cells are more recently added).\label{fig:Cu+MC}}
\end{figure}

While Ni(111) provides a convenient system for atomistic analysis,
symmetry-breaking effects are equally important for other substrates
without perfect epitaxy with graphene, the foremost being copper.
\textbf{Fig. \ref{fig:Cu+MC} (a)} presents a scanning electron microscope
image of graphene islands on a Cu foil. The growth was carried out
in a tube furnace CVD system, similar to previous work \cite{2013haotherole}.
For this sample we used an oxygen-free Cu substrate with 0.1 torr
H\textsubscript{2} pressure and $10^{-3}$ torr methane pressure.
The growth temperature was 1035 \textdegree C and the growth time
was 20 min. Several Cu grains are seen, with many graphene islands
(dark) on each. Even though all graphene islands grew simultaneously
at the same conditions, we see two distinct shape classes. Nearly
all islands are hexagonal, but some are almost perfect while others
are elongated. All islands on a single Cu grain belong to the same
class (except for cases with several islands colliding within the
same grain or across the boundaries %
\footnote{After a collision within the same Cu grain the shape of composite
island will approach the original kinematic Wulff construction as
it grows larger compared to the separation between the original two.
Thus one could still grow large single-crystals with well-defined
shape even starting from more than one nucleus.%
}). Furthermore, graphene islands on each grain are aligned, which
is especially noticeable for elongated islands. The density of islands
and their size is similar between the grains. This suggests that the
basic underlying mechanisms in their growth are the same, and the
shape difference is determined by subtle differences between crystallographic
surfaces of Cu. Indeed, electron back-scatter diffraction studies
established that hexagonal domains form on Cu(111), while elongated
domains grow on Cu (100) or (110) \cite{2013haotherole,2013murdockcontrolling}.

To understand the shapes in \textbf{Fig. \ref{fig:Cu+MC} (a)} we
consider the symmetry of the graphene+Cu system. Though the (111)
surface is triangular, translational incommensurability of the lattices
means that a growing edge will have a different positioning with respect
to the underlying Cu atoms at different times, and on average, the
symmetry-breaking effects will compensate. Interestingly, this precludes
the sliding mode of symmetry breaking (as in Ni(111)) and restores
the $C_{6}$ symmetry for the composite system. For rectangular (110)
and square (100) surfaces, two parallel graphene edges should align
with one of the orthogonal basic crystallographic directions of the
surface, but the other four edges will remain misaligned with the
substrate. Thus the six edges will be split in two sets of two and
four (unlike three and three on Ni). During kink-flow growth, incommensurability
of edges with the underlying substrate will again be averaged out,
but this time the averaging will differ between the families. Again,
the rotational symmetry of graphene islands is reduced to the common
divisor of 6 and 2 or 4, respectively, i.e., $C_{2}$.

Building on the understanding of how the substrate modulates the growth
velocities in different directions one can model this and other possibilities
using coarse-grained MC simulation as follows. Graphene is represented
by a triangular lattice (nodes are hexagon centers), starting with
a single occupied point. At each time step, all vacant cells with
occupied neighbors are classified either as \emph{Z} sites (three
occupied neighbors along a straight line) or \emph{k} (kink; armchair
edge is an `array of kinks' \cite{2009dingdislocation}). The relative
probability $P$ of addition to a \emph{Z} site (nucleation of a new
atomic row, $P\equiv P_{Z}/P_{k}\sim\exp\nicefrac{\left(E_{k}-E_{Z}\right)}{k_{\text{B}}T}$
in the atomistic calculations) is the input parameter. Probabilities
of all sites form a distribution which is sampled to determine the
next site to be occupied, and the process is iterated. As a result
we obtain the shape in \textbf{Fig. \ref{fig:Cu+MC} (b)}, which is
the familiar graphene hexagon frequently observed on Cu(111) \cite{2011yucontrol,2013haotherole},
liquid Cu \cite{2012genguniform}, and in \textbf{Fig. \ref{fig:Cu+MC}
(a)}, in agreement with the nanoreactor model predictions for isotropic
substrates \cite{2012artyukhovequilibrium}.

Recalling our analysis of growth kinetics on Ni (\textbf{Fig. 2}),
the essential physical insight that explains triangular growth shapes
is the difference between probabilities to initiate a new row on \emph{Z}
vs. \emph{K} edges. Our MC model can naturally capture this with two
independent parameters for two inequivalent crystallographic orientations
of \emph{Z} edges, $P_{\nabla}$ and $P_{\Delta}$. \textbf{Fig. \ref{fig:Cu+MC}
(c)} shows the sharp triangle from a run with $P_{\Delta}/P_{\nabla}=10^{-3}$,
corresponding to our first-principles results for $\Delta E$. Only
\emph{$\Delta$} edges are present, and it is in perfect agreement
with the analytical kinematic Wulff shape of \textbf{Fig. \ref{fig:Wulff}}.
\textbf{Fig. \ref{fig:Cu+MC} (d)} shows the truncated triangle produced
in a run with $P_{\nabla}/P_{\Delta}=10^{-1}$, barely showing any
\emph{$\Delta$} edge fragments. Thus, six-sided shapes are only possible
when the growth barrier difference is small, $\left|E_{\nabla}-E_{\Delta}\right|\lesssim2\text{--}3\, k_{\text{B}}T$,
or no more than 0.2 eV for typical graphene CVD conditions, which
is a rather close coincidence. 

For rectangular surfaces such as Cu(110) or Ge(110) \cite{2014leewaferscale},
or twofold-symmetric stackings on Ni(111), two input probabilities
are again needed, now for the two `horizontal' and four `diagonal'
directions, $P_{=}$ and $P_{\langle\,\rangle}$. Typically one would
expect the edges that are aligned with close-packed surface `grooves'
to grow slower, resulting in $P_{=}<P_{\langle\,\rangle}$. This produces
elongated shapes such as \textbf{Fig. \ref{fig:Cu+MC} (e)}, closely
resembling the high aspect ratio islands in \textbf{Fig. \ref{fig:Cu+MC}
(a)}. Cu(100) is similar, but there are now two orthogonal close-packed
directions for long graphene edges to align with. This will produce
two rather than one preferred alignments at a 90\textdegree angle
with each other within the same Cu grain (\textbf{Fig. \ref{fig:Cu+MC}
(f)}) as observed experimentally \cite{2013murdockcontrolling,2011ogawadomainstructure}.
Finally, if $P_{=}>P_{\langle\,\rangle}$, the shape shown in \textbf{Fig.
\ref{fig:Cu+MC} (g)} results.

It is remarkable how a simple MC model informed by atomistics allows
a unified description of Ni(111), all surfaces of Cu (including liquid),
and pretty much any metal surface without an epitaxial match with
graphene just based on its symmetry. It can similarly be applied to
model any other graphene-like material with inequivalent sublattices,
such as BN \cite{2012ismachtowardthe,2012kimsynthesis} or transition
metal dichalcogenides. By the same token, growth units larger than
hexagons \cite{2013dongkinetics} can be treated. Going even further
one can emulate diffusion in this model. This is achieved by making
the growth probabilities depend not only on site type, but also on
the number of unoccupied cells within some distance. Edges of protrusions
have better access to feedstock supply at the free catalyst surface,
producing diffusion instabilities. This refinement reproduces sawtooth
patterns seen on the edges of metal chalcogenide islands \cite{2013vanderzandegrainsand,2013zhangcontrolled}
with a characteristic dendritic but not finger-like morphology (\textbf{Fig.
\ref{fig:Cu+MC} (h)}), reminiscent of the Sierpinski fractal.

\medskip

In summary, the symmetry of emergent carbon islands reflects not the
symmetry of graphene per se but rather the combined symmetry of its
stacking on a substrate surface, which generally is lower than either
graphene (hexagonal) or the surface (hexagonal, square, rectangular\ldots{}).
On epitaxially matched surfaces such as Ni(111) or Co(0001) the symmetry
breaking effect is particularly apparent at the edges, resulting in
different ground-state structures (\emph{Z} vs. Klein) for different
directions ($\nabla,\,\Delta$), and causing equilibrium shapes with
a (typically, mild) violation of inversion symmetry. However, in kinetics,
the symmetry-lowering interactions become exponentially amplified
as $\sim\exp\left(-E/k_{\text{B}}T\right)$, and Klein edges grow
much faster than \emph{Z}, resulting in triangular growth shapes with
only \emph{Z} edges. Similarly, exponentiation can make symmetry effects
strongly pronounced in nucleation, so that edge energy differences
can play a decisive role in selection of the graphene--Ni(111) stacking.
We apply this insight to growth on Cu, where different graphene island
morphologies are concurrently observed on different crystalline grains
of the same foil, using a Monte Carlo growth model that draws upon
our Ni(111) analysis but can be tuned to any substrate symmetry, commensurate
or incommensurate with graphene, crystalline or liquid. Since crystal
symmetry of the substrate dictates both the shape of islands and their
alignment, single-crystalline substrates offer better control over
both the morphology of graphene islands and grain boundaries in the
resulting films. This improved understanding of the role of substrate
symmetry in graphene growth is crucial for improving the quality \cite{2014tetlowgrowthof}
or engineering grain boundaries \cite{2014yazyevpolycrystalline}
in CVD graphene.

\bibliographystyle{h-physrev}
\bibliography{triangles}

\begin{thebibliography}{10}

\bibitem{2004novoselovelectric}
K.~S. Novoselov {\em et~al.},
\newblock Science {\bf 306}, 666 (2004).

\bibitem{2011colemantwodimensional}
J.~Coleman {\em et~al.},
\newblock Science {\bf 331}, 568 (2011).

\bibitem{2008bolotinultrahigh}
K.~Bolotin {\em et~al.},
\newblock Solid State Commun. {\bf 146}, 351 (2008).

\bibitem{2014tetlowgrowthof}
H.~Tetlow {\em et~al.},
\newblock Phys. Rep. {\bf 542}, 195 (2014).

\bibitem{2014yazyevpolycrystalline}
O.~V. Yazyev and Y.~P. Chen,
\newblock Nat. Nanotech.  (2014).

\bibitem{2012artyukhovequilibrium}
V.~I. Artyukhov, Y.~Liu, and B.~I. Yakobson,
\newblock Proc. Natl. Acad. Sci. U.S.A. {\bf 109}, 15136 (2012).

\bibitem{2011yucontrol}
Q.~Yu {\em et~al.},
\newblock Nat. Mater. {\bf 10}, 443 (2011).

\bibitem{2012genguniform}
D.~Geng {\em et~al.},
\newblock Proc. Natl. Acad. Sci. U.S.A. {\bf 109}, 7992 (2012).

\bibitem{2013haotherole}
Y.~Hao {\em et~al.},
\newblock Science {\bf 342}, 720 (2013).

\bibitem{2013maedgecontrolled}
T.~Ma {\em et~al.},
\newblock Proc. Natl. Acad. Sci. U.S.A. {\bf 110}, 20386 (2013).

\bibitem{2014chennearequilibrium}
J.~Chen {\em et~al.},
\newblock Adv. Mater. {\bf 26}, 1348 (2014).

\bibitem{2011liubnwhite}
Y.~Liu, S.~Bhowmick, and B.~I. Yakobson,
\newblock Nano Lett. {\bf 11}, 3113 (2011).

\bibitem{2010liugraphene}
Y.~Liu, A.~Dobrinsky, and B.~I. Yakobson,
\newblock Phys. Rev. Lett. {\bf 105}, 235502 (2010).

\bibitem{1993kresseabinitio}
G.~Kresse and J.~Hafner,
\newblock Phys. Rev. B {\bf 47}, 558 (1993).

\bibitem{1996kresseefficient}
G.~Kresse and J.~Furthm{\"u}ller,
\newblock Phys. Rev. B {\bf 54}, 11169 (1996).

\bibitem{1994blochlprojector}
P.~E. Bl{\"o}chl,
\newblock Phys. Rev. B {\bf 50}, 17953 (1994).

\bibitem{1999kressefromultrasoft}
G.~Kresse and D.~Joubert,
\newblock Phys. Rev. B {\bf 59}, 1758 (1999).

\bibitem{1980ceperleygroundstate}
D.~M. Ceperley and B.~J. Alder,
\newblock Phys. Rev. Lett. {\bf 45}, 566 (1980).

\bibitem{1996perdewgeneralized}
J.~P. Perdew, K.~Burke, and M.~Ernzerhof,
\newblock Phys. Rev. Lett. {\bf 77}, 3865 (1996).

\bibitem{1997perdewgeneralized}
J.~P. Perdew, K.~Burke, and M.~Ernzerhof,
\newblock Phys. Rev. Lett. {\bf 78}, 1396 (1997).

\bibitem{Note1}
See Supplemental Material at http://prl.aps.org for for details of computations
  and edge energy fitting.

\bibitem{1994kleingraphitic}
D.~Klein,
\newblock Chem. Phys. Lett. {\bf 217}, 261 (1994).

\bibitem{2013wagnerstablehydrogenated}
P.~Wagner {\em et~al.},
\newblock Phys. Rev. B {\bf 88}, 094106 (2013).

\bibitem{2013paterainsitu}
L.~L. Patera {\em et~al.},
\newblock {ACS} Nano {\bf 7}, 7901 (2013).

\bibitem{1951herringsometheorems}
C.~Herring,
\newblock Phys. Rev. {\bf 82}, 87 (1951).

\bibitem{2014garcia-lekuespindependent}
A.~Garcia-Lekue {\em et~al.},
\newblock Phys. Rev. Lett. {\bf 112}, 066802 (2014).

\bibitem{2012olleyieldand}
M.~Olle, G.~Ceballos, D.~Serrate, and P.~Gambardella,
\newblock Nano Lett. {\bf 12}, 4431 (2012).

\bibitem{2014artyukhovwhynanotubes}
V.~I. Artyukhov, E.~S. Penev, and B.~I. Yakobson,
\newblock Nat. Commun. {\bf 5}, 4892 (2014).

\bibitem{2013liequilibrium}
M.~Li {\em et~al.},
\newblock Phys. Rev. B {\bf 88}, 041402 (2013).

\bibitem{2010lahirianextended}
J.~Lahiri, Y.~Lin, P.~Bozkurt, I.~Oleynik, and M.~Batzill,
\newblock Nat. Nanotech. {\bf 5}, 326 (2010).

\bibitem{2014bianchiniatomicscale}
F.~Bianchini, L.~L. Patera, M.~Peressi, C.~Africh, and G.~Comelli,
\newblock J. Phys. Chem. Lett. , 467 (2014).

\bibitem{Note2}
After a collision within the same Cu grain the shape of composite island will
  approach the original kinematic Wulff construction as it grows larger
  compared to the separation between the original two. Thus one could still
  grow large single-crystals with well-defined shape even starting from more
  than one nucleus.

\bibitem{2013murdockcontrolling}
A.~T. Murdock {\em et~al.},
\newblock {ACS} Nano {\bf 7}, 1351 (2013).

\bibitem{2009dingdislocation}
F.~Ding, A.~R. Harutyunyan, and B.~I. Yakobson,
\newblock Proc. Natl. Acad. Sci. U.S.A. {\bf 106}, 2506 (2009).

\bibitem{2014leewaferscale}
J.-H. Lee {\em et~al.},
\newblock Science {\bf 344}, 286 (2014).

\bibitem{2011ogawadomainstructure}
Y.~Ogawa {\em et~al.},
\newblock J. Phys. Chem. Lett. {\bf 3}, 219 (2011).

\bibitem{2012ismachtowardthe}
A.~Ismach {\em et~al.},
\newblock {ACS} Nano {\bf 6}, 6378 (2012).

\bibitem{2012kimsynthesis}
K.~K. Kim {\em et~al.},
\newblock Nano Lett. {\bf 12}, 161 (2012).

\bibitem{2013dongkinetics}
G.~Dong and J.~W.~M. Frenken,
\newblock {ACS} Nano {\bf 7}, 7028 (2013).

\bibitem{2013vanderzandegrainsand}
A.~M. van~der Zande {\em et~al.},
\newblock Nat. Mater. {\bf 12}, 554 (2013).

\bibitem{2013zhangcontrolled}
Y.~Zhang {\em et~al.},
\newblock {ACS} Nano {\bf 7}, 8963 (2013).

\end{thebibliography}

\end{document}